\documentclass[10pt,journal,compsoc]{IEEEtran}
\usepackage{amssymb}

\usepackage{amsmath,amssymb,amsfonts}
\usepackage{algorithmic}
\usepackage{algorithm}
\usepackage{comment}
\usepackage{rotating}
\usepackage{tabularx}
\usepackage{lscape} 
\usepackage{subfig}
\usepackage{adjustbox}
\usepackage{multirow}
\usepackage{url}
\usepackage{booktabs}
\usepackage[utf8]{inputenc}
\usepackage{float}
\usepackage{graphicx}

\ifCLASSOPTIONcompsoc
  \usepackage[nocompress]{cite}
\else
  \usepackage{cite}
\fi
\ifCLASSINFOpdf
\else
\fi
\begin{document}
%
\title{SDP: Scalable Real-time Dynamic Graph Partitioner}
%
%
%
%

\author{Md Anwarul Kaium Patwary,~\IEEEmembership{Member,~IEEE,}
        Saurabh Garg,~\IEEEmembership{Member,~IEEE,} Sudheer Kumar Battula
        and~Byeong Kang,~\IEEEmembership{Member,~IEEE}

}

%
%

\markboth{IEEE Transactions on Services Computing ,~Vol.~, No.~, }%
{Shell \MakeLowercase{\textit{et al.}}: Bare Demo of IEEEtran.cls for Computer Society Journals}
%



\IEEEtitleabstractindextext{%
\begin{abstract}
Time-evolving large graph has received attention due to their participation in real-world applications such as social networks and PageRank calculation. It is necessary to partition a large-scale dynamic graph in a streaming manner to overcome the memory bottleneck while partitioning the computational load. Reducing network communication and balancing the load between the partitions are the criteria to achieve effective run-time performance in graph partitioning. Moreover, an optimal resource allocation is needed to utilise the resources while catering the graph streams into the partitions. A number of existing partitioning algorithms (ADP, LogGP and LEOPARD) have been proposed to address the above problem. However, these partitioning methods are incapable of scaling the resources and handling the stream of data in real-time.

In this study, we propose a dynamic graph partitioning method called Scalable Dynamic Graph Partitioner (SDP) using the streaming partitioning technique. The SDP contributes a novel vertex assigning method, communication-aware balancing method, and a scaling technique to produce an efficient dynamic graph partitioner. Experiment results show that the proposed method achieves up to 90\% reduction of communication cost and 60\%-70\% balancing the load dynamically, compared with previous algorithms. Moreover, the proposed algorithm significantly reduces the execution time during partitioning.
\end{abstract}

}

\maketitle

\IEEEdisplaynontitleabstractindextext

%
\IEEEpeerreviewmaketitle

\IEEEraisesectionheading{\section{Introduction}\label{sec:introduction}}

\IEEEPARstart{I}{n} recent days most of the graph-oriented applications have a dynamic behaviour, which means that the vertex or edge might go off or gain a new vertex or edges. For example, thousands of Twitter users update their Tweets per seconds \cite{Vaquero:2013:APL:2523616.2525943}. This behaviour of the dynamic graphs creates the computational load imbalance between partition and increases the edge-cuts and communication cost as well.

Most real-world graph applications such as social networks, weather forecast tend to receive graph data continuously as a stream of graph data in a real-time manner. It is necessary to have such a graph partitioning algorithm that can distribute the stream data among the partitions in a one-pass manner as the vertex arrives. A streaming graph partitioning algorithm receives the vertices one by one and decides the respective partitions with little connectivity information of a vertex. In one-pass manner the vertex can only be seen once before assigning to respective partitions. When a graph gets updated over time, it is necessary to keep the computational load balanced, keeping the communication to a minimum. If the algorithm has to visit all the partitions again and revisit the whole graph to perform the repartitioning for an updated graph, it is very expensive in terms of computational time.

The computational load of a partition in a dynamic graph always changes over time, by adding or removing vertex elements from a partitioned graph. Since in streaming partitioning the incoming number of vertices is unknown and the number of vertices might be removed anytime from a partition, a huge imbalance between partitions is created. Consequently, this creates more cut edges and causes unbalanced partitions. Unbalanced partitions might also cause an unnecessary allocation of a partition, which is a waste of computational resources. A number of algorithms are proposed \cite{Vaquero:2013:APL:2523616.2525943, Walshaw:1997:PDG:281659.281661,6676492} to address the load balanced issue for dynamic graph. Of them no algorithms used the streaming graph partitioning technique. An adaptive partitioning method \cite{Vaquero:2013:APL:2523616.2525943} was proposed for dynamic graphs. The main idea of this technique is to migrate the vertices/edges from one partition to another meeting some criteria towards reducing the load imbalance and communication cost. However, this technique has a huge communication overhead when migrating vertices.    

In order to cater for the ever-increasing computational load as per the demands of an application, scalability becomes an important factor in dynamic graph partitioning. In this study, we also propose a dynamic machine allocation method to allocate a new machine as per the demands of the computational load. Dynamic allocation of a machine is another important aspect in balanced graph partitioning, as over time the size of a graph continuously changes. It is important to consider a flexible allocation of a new partition according to the computational load. This study addresses these features by allowing the decrease and increase of the number of partition allocations, according to the computational load. We use a capacity threshold to decide the allocation of a new machine or shutting down of an unused machine from the Cloud.        

A communication aware balancing strategy is also taken into account when assigning vertices to a corresponding partition. It trades-off with the number of communications, while minimising the load imbalance between partitions. 

The main contributions of this research are as follows,

\begin{enumerate}
	\item A dynamic partitioning algorithm which can handle the dynamic changes of a large-graph when new vertices and edges are added or removed continuously over time, and which can assign vertices and edges to an appropriate machine using a novel vertex assigning technique.
	\item  A dynamic partitioning algorithm which accepts graph data in a streaming manner.
	\item  A communication-aware balancing strategy used dynamically to reduce the imbalance of computational load among the partitions.   
	\item A vertex migration technique, in order to scale up or down the resources to optimise the resources and minimise the resource cost.
\end{enumerate}


The rest of the paper is organised as follows. We discuss the dynamic graph partitioning problem in Section \ref{problem-statement}. In Section \ref{related-work}, we discuss the related work and the most recent advancements in dynamic graph partitioning. We discuss the system architecture and proposed dynamic graph partitioning algorithm in Section \ref{system-arch-section}. In Section \ref{experiment-section}, the datasets, experimental scenario, performance matrices and evaluation setup are discussed. In Section \ref{resul-discussion}, results for this study are analysed. Finally, we discuss our conclusion and future work of this study in Section \ref{conclusion}.




\section{Problem Statement}
\label{problem-statement}
\begin{figure}[h]
	\includegraphics[width=\linewidth]{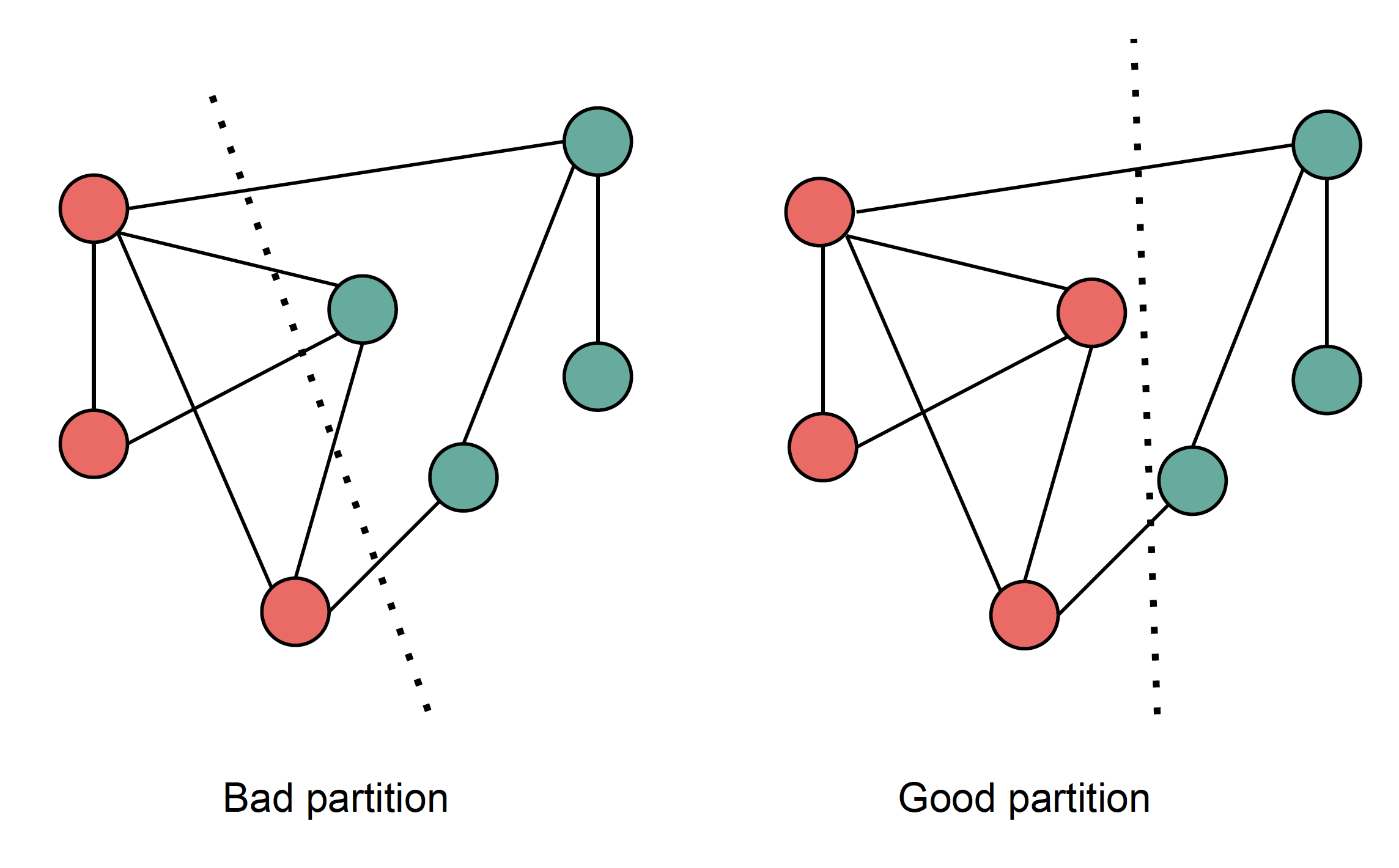}
	\caption{Graph partition}
	\label{partition_fig}
\end{figure}

Partitioning a graph has two main objectives, minimising the cut edges and balancing the load between partitions. This partitioning is done by distributing vertices and their edges to the machines in a distributed system. Figure \ref{partition_fig} shows the example of bad and good partition. The more cut edges between partitions creates a bad partitioning in a distributed system.

\newtheorem{Problem}{Problem}
\begin{Problem}
	Partitioning of a dynamic graph $G=(V,E)$ into $k$ number of subgraphs  and allocate each subgraphs to the $kth$ partition. The number of vertices $V$ in each partition increases or decreases over time $t$, so the number of vertices of a partition after $t$ time would be $|V_k(t)|$. The graph partitioning technique always aim to reduce the cut edges $E(u,v)$ between partitions. When two different end points ($u$ and $v$) of an edge $E$ resides in different partitions, such that, $min E(u,v) = \sum_{i=1}^{k} E(u,v)$. 
\end{Problem}

\begin{Problem}
	In $k$ way graph partitioning, the algorithm always tends to divide the entire graph $G$ into $k$ number of sub-graphs. In a dynamic graph partitioning the size of the graph is continuously growing and the number of partition is dynamic and unknown. The number of partitions $P_k$ should be allocated according to the computational load over time $t$ as follows: $|P_k(t)|$, such that the $k$ value increases or decreases as per the computational load.     	
\end{Problem}

\section{Related Work}
\label{related-work}
Various dynamic graph partitioning technique have been developed. In order to overcome the computational complexity using the traditional balanced graph-partitioning technique a number of large-scale of streaming graph partitioning proposed in ~\cite{Stanton2012},\cite{Tsourakakis:2014:FSG:2556195.2556213}, \cite{10.1145/3290688.3290711} and \cite{Cheng:2012:KTP:2168836.2168846}. These techniques managed to reduce the run-time during partitioning. Linear Deterministic Greedy (LDG)\cite{Stanton2012} heuristic assigns a vertex to the partition with which it shares the most edges. This streaming partitioning method makes heuristics scalable in the size and the number of partitions of graphs. It has significant speed up achievement of PageRank computations on Spark\cite{Zaharia2010} by 18\% to 39\% for large social networks. FENNEL leverages modularity maximisation \cite{Newman06062006} to deploy a greedy strategy for maintaining balanced partitions. It has also improved performance regarding the communication cost and run-time, while computing iteratively graph data in a distributed system. Another greedy heuristic algorithm proposed in\cite{Ahmed:2013:DLN:2488388.2488393} uses an unweighted, deterministic greedy algorithm, instead of using the weighted penalty function in order to partition vertices. This algorithm also uses a factorisation technique that aims to reduce the neighbouring vertices rather than the edges across the partition. In other words, a vertex-cut partitioning technique is employed here, which is well-suited for large-scale natural graphs. However, the proposed techniques has not explored the issue in removing and adding data while processing the graph in streaming manner which is most demanding aspect in today's graph processing system


An adaptive dynamic partitioning, called xDGP \cite{DBLP:journals/corr/VaqueroCLM13} was developed to improve the graph partitioning  performance. It uses an iterative vertex migration algorithm that relies on local information only. It has been demonstrated that a significant improvement was achieved in graph partitioning, reducing execution time by more than 50\%. It also adapts the graph partitioning structure by balancing load with a large number of changes. Another adaptive unstructured meshes dynamic partitioning algorithm \cite{Walshaw:1997:PDG:281659.281661} was proposed with parallelisation. This algorithm uses a relative gain optimisation technique, which aims to balance workload and reduce the inter-partition communication overhead. A few series of adaptive refined meshes were applied for the purpose of the experiment, and the results indicate that they provide better partitioning than a static partitioner. However, the proposed methods unable to handle the real-time graph data.

 
Vertex replication is another technique to handle the ever-changing graph in a distributed environment. Vertex replication imitates the vertex in a partition to reduce the communication cost in a distributed graph processing system. A vertex replication algorithm \cite{Huang:2016:LLE:2904483.2904486} was proposed with the aim of attaining better access locality of a vertex, by replicating the vertex which resides in another partition. Eventually, it does minimise the communication cost across the network.  

\vspace{-1pt}
A few more researchers proposed the vertex replication method in graph partitioning, while minimising the workload imbalance and inter-machine communication in a distributed network. Of them, dynamic replication-based partitioning was proposed in \cite{Yang:2012:TEP:2213836.2213895}, and this replicates the vertex adaptively, based on the change of workload. To improve performance during the frequent changes in workload, an historical log-based partitioning technique called LogGP was proposed  \cite{Xu:2014:LLD:2733085.2733097}. LogGP framework analyses and reuses the historical statistical information to refine the partitioning result. It has great advantages in utilising the historical partitioning results to generate a hypergraph. The authors argue that running statistical analysis of historical partitioning logs can provide an improvement on partitioning results. Alleviation load skew at query time is another benefit of vertices replication after distributing the large graph-structured network. In \cite{Duong:2013:SSN:2433396.2433424} this replication feature is presented. If there is no replication, popular nodes become overwhelmed by request, in a partition for the value of those nodes. 

Dynamic graphs sometimes require a repartitioning process to maintain the balance of graph-partitioned data in order to improve system performance. Good partitioning algorithms with repartitioning features are in demand for handling huge dynamic graph data. A study was undertaken on repartitioning online social network data in \cite{6172626}. The authors aimed to improve the scalability by reducing the inter-partitioning communication. A replication method was used to reduce the communication among nodes. An in-memory based dynamic partitioning technique was proposed in \cite{Mondal:2012:MLD:2213836.2213854} to handle the large dynamic graph. This algorithm achieved significant low-latency communication in query processing. The authors provided a vertex replication policy that monitors the incoming vertices and decides what data to replicate. It was evaluated on a social network graph, and the result shows that this technique reduced the network bandwidth significantly. Moreover, the technique also handled a very large graph efficiently. Proper placement of a newly added vertex, in a dynamic graph, by using the cost-effective method, was proposed in \cite{Abdolrashidi7584916}. A vertex migration technique was also added to this study in order to balance the partitions, due to deletion of vertices from a partition. The migration of the vertex depends on the latency and communication cost of the particular vertex being migrated. The authors proposed a set of heuristics to reduce communication cost, and to balance the partitions. However, these heuristics do not accept the graph data in a stream manner and do not make any decisions in real time. A hash based dynamic graph partitioning proposed \cite{WANG2019804}. The proposed technique analysed the graph nodes locality of local machine before partitioning the nodes. This dynamic partitioning technique does not cover the resource auto-scaling.    

\begin{table}[H]
	\centering
	\caption{Summarization of dynamic graph partitioning}
	\label{related-work-table}
	\setlength{\tabcolsep}{3pt}
	\begin{adjustbox}{width=\linewidth} 
		\begin{tabular}{ |p{3cm}|p{1.5cm}|p{1.5cm}|p{1.5cm}|p{2.0cm}|  }
			\hline
			\multirow{2}{*}{Algorithm}{\textbf{}} & \multicolumn{3}{c|}{Features}{\textbf{}}\\ 
			\cline{2-4}
			& Auto-scalling & Distributed & Load Balancing\\
			\hline
			Linear Deterministic Greedy(LDG)\cite{Stanton2012}  &No & Yes &No \\
			\hline
			Natural Graph Factorization\cite{Ahmed:2013:DLN:2488388.2488393}&No &Yes  &No \\
			\hline
			LOOM\cite{firth2016workload} &No &Yes  &No \\
			\hline
			STINGER\cite{Riedy:2013:MSD:2425676.2425689}&	No&	No&	No\\
			\hline
			Planted Partition\cite{Tsourakakis:2015:SGP:2817946.2817950}&	No&	No&	No\\
			\hline
			HDRF\cite{Petroni:2015:HSP:2806416.2806424}&	No&	Yes&	No\\
			\hline
			HoVerCut\cite{7584914}&	No&	Yes&	Yes\\
			\hline
			Vertex Migration\cite{Abdolrashidi7584916}&Yes&Yes&No\\
			\hline
			TSH\cite{WANG2019804}&No&Yes&No\\
			\hline
			\textbf{SDP}&Yes&Yes&Yes\\
			\hline
		\end{tabular}
	\end{adjustbox}
\end{table}

\begin{algorithm*}
	\begin{flushleft}
		\textbf{INPUT:} $V$ = set of partitioned vertices, $P$ = number of partition indexes,  $v$ = vertex arrived in stream, $edge<v_1,v_2>$= edge arrived in the stream, $\alpha$= type of input, $E(v)$ is the associated edges arrived with vertex $v$, $partitionInfoMap<p<List>>$, $edgeInfoMap[]<vertex, List<edges>>$, $TH$ = balancing threshold.
		$averageLoad$ = average load of all the partitions. 	 
	\end{flushleft}
	\begin{algorithmic}[1]
		\STATE $MAXCAP \gets$ the maximum capacity of each partition
		\IF{($ \alpha = add$)}
		\STATE $thresHold \gets addingThreshold(|E|,|P|) $
		\IF{($MAXCAP \leq thresHold$)}
		\STATE $updateSummery(P+1, v, MAXCAP, averageLoad)$
		\ENDIF
		\IF{($P > 1$)}
		\STATE $\sigma \gets findImbalance(P, partitionInfoMap<p,<List>>)$
		\ENDIF
		\IF{($\sigma > TH$)}
		\STATE $partitionIndex \gets assignVertex(v,P,V, E(v))$  \COMMENT{create a new partition and assign the vertices to the partition.}
		\STATE $updateSummary (partitionIndex, v, MAXCAP, averageLoad)$ \COMMENT{update the partition index in graph summary}
		\ELSE 
		\STATE $partitionIndex \gets findMinimum(partitionInfoMap<p<List>>,P)$
		\COMMENT{If the load difference is less than the threshold then find the minimum load partition}
		\STATE $assignVertex(v,partitionIndex,V, E(v))$  \COMMENT{ assign the vertices to the  partition which has minimum load.}
		\STATE $updateSummary (partitionIndex, v, MAXCAP, averageLoad)$
		\ENDIF
		\ELSE
		\IF{($ \alpha = deleteVertex$)}
		\STATE $deleteVertex (v, P, edgeInfoMap[] <vertex, List<edges>>, partitionInfoMap<p<List>>) $
		\COMMENT{Delete the vertices $v$ and their associated edges.}
		\STATE $updateSummary (partitionIndex, v, MAXCAP, averageLoad)$
		\ELSE
		\IF{($ \alpha = deleteEdge$)}
		\STATE $deleteEdges(edge, <v_1, v_2>, P, edgeInfoMap[]<vertex, List<edges>>)$
		\COMMENT{Delete the edges $edge$.}
		\ENDIF
		\ENDIF
		\ENDIF
	\end{algorithmic}
	\caption{Dynamic Partitioning }
	\label{main-algo}
\end{algorithm*}

\section{SDP Architecture and Algorithm}
\label{system-arch-section} 

This section describes the complete architecture and algorithm of our proposed dynamic partitioning technique. 

\subsection{SDP Architecture}

\begin{figure}
	\includegraphics[width=\linewidth]{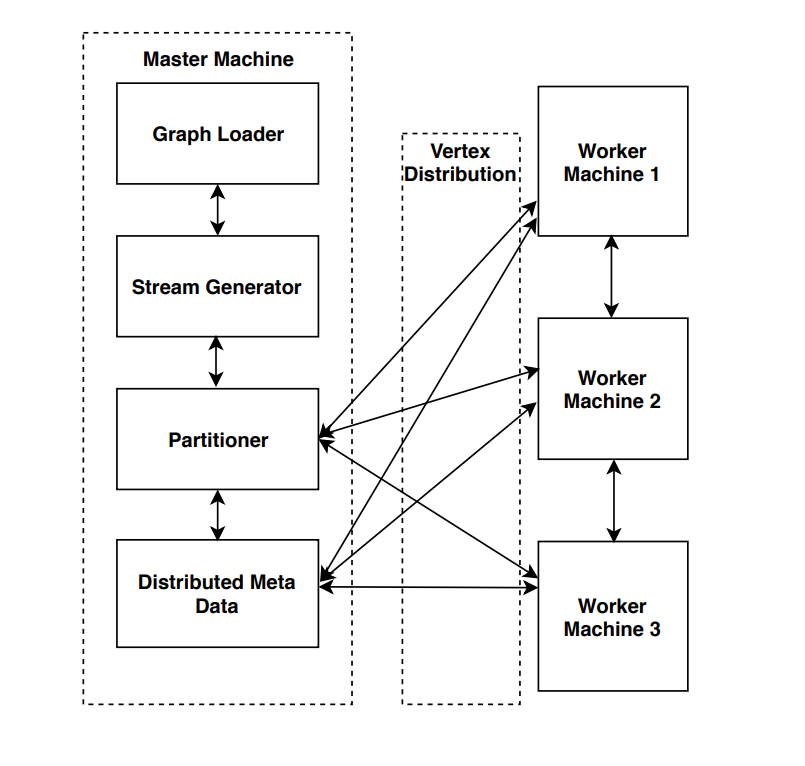}
	\caption{System Architecture}
	\label{System_arch_fig}
\end{figure}

The graph data distributes to the number of machines in the distributed system in order to balance the computational load evenly between the machines. A Graph Loader also resides in the master machine which decides the nature of the input. The master machine takes the input, and a stream generator resides in the master machine to generate the stream of data from the Graph Loader. The stream generator forwards the input to the partitioner to perform the addition/deletion. Our dynamic partitioning method accepts three kinds of inputs(for example add, delete a vertex, and delete edge). Figure \ref{System_arch_fig} shows the architecture of this study.

The partitioning process starts with one worker machine and adds the partition dynamically, according to the load. The adding criteria of new partitions is explained in Section \ref{scalability}.

The components of system architecture are as follows:

\begin{itemize}
	\item \textbf{Graph Loader:} The Graph Loader loads the input from the disk memory. For example: add vertex, delete vertex and delete edge. The loader receives input from the disk uniformly and at random for this purpose and forwards to the Stream Generator in order to create the stream of data before forwarding to the partitioner. 
	
	\item \textbf{Stream Generator:} A stream generator resides in the master machine to generate the stream of graph data from input dataset. Each vertex arrives with its associated edges in the stream, in sequentially from the Graph Loader.  Stream generator loads the entire graph dataset and sends the graph with multiple threads to the graph loader queue. Graph loader reads the data from the queue in parallel manner. It is responsible for forwarding the graph input to the partitioning algorithm for the purpose of adding or deleting. 
	
	\item \textbf{Distributed Meta Data:} It is located in the master machine to store the graph information such as vertices, number of edges and worker IP, which can be used for partitioning purposes of the upcoming graph data by the partitioner.  
	
	\item \textbf{Data Receiver in Worker Machine:} A data receiver resides in the worker machines to receive the vertices and edges from the master machine and send the acknowledgement to the master to update the metadata information. 
\end{itemize}

\subsection{SDP: Dynamic Graph Partitioner}
\label{Dynamic-Graph}
The algorithm aims to minimise the edge-cut among partitions and reduce the partition imbalance as low as possible. Algorithm \ref{main-algo} represents the entire dynamic graph partitioning process. The SDP algorithm receives the graph data in streaming manner which means each time a vertex arrives for partitioning, the algorithm decides a suitable partition to allocate that vertex immediately. The algorithm stores the summary of partitioning results, in a distributed meta data file in the master machine. The meta data is used as a reference to allocate the future vertices. The summary contains vertex information and its allocated partition index. After allocating or deleting each vertex or edges from a partition, the graph summary will be updated accordingly. Algorithm \ref{update:summary:algo} depicts the updating graph summary.



\begin{figure}
	\includegraphics[width=\linewidth]{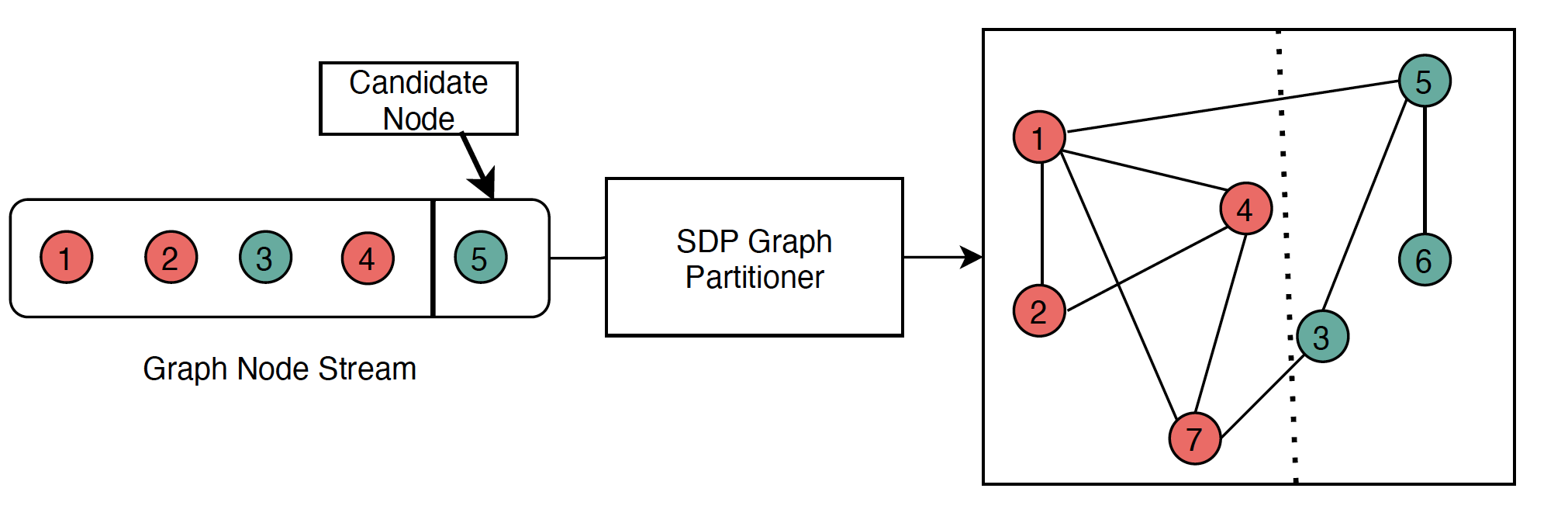}
	\caption{Graph Stream in Partitioning}
	\label{Stream_fig}
\end{figure}

\begin{algorithm}
	\begin{flushleft}
		\textbf{INPUT:} $p$ = partition index, $v$ = arrived vertex, $MAXCAP$ = maximum capacity of a partition, $averageLoad$ = the average load of the partitions  \\
	\end{flushleft}
	\begin{algorithmic}
    \STATE Create partitionInfoMap with the partition index as key and list of vertices.

\IF{$p \; \exists! \; partitionInfoMap$}
\STATE Get the list of vertices $VertexList<v>$ associated with the partition index $p$. 
\STATE Get the list of vertices associated with the partition index $p$. 
\STATE Add the current arrived vertec $v$ to the vertex list $VertexList<v>$.
\ELSE
\STATE Create a new entry with the partition index $p$ and arrived vertex $v$,
		\ENDIF  
	\end{algorithmic}
	\caption{Update the graph summary}
	\label{update:summary:algo}
\end{algorithm}
where, $parttionInfoMap$ is to store the summary of the graph and the partition index. The partitioning algorithm uses this information to assign a vertex to a proper location. The algorithm starts by taking a tuple $V<vertex, edges>$ as an input, and the input comes in a single pass manner sequentially. It receives tuple at any point of time as a stream of data is added to a machine and is removed over time. Based on the type of input it receives, the algorithm acts accordingly. The type of input is decided by the Graph Loader in the master machine.  
If the algorithm receives an input to add the vertices, the vertex allocation technique is employed to assign a vertex to a partition. Vertex allocation technique is described in Section \ref{vertex-assigning-section} and the assigning algorithm is depicted in Algorithm \ref{Vertex-Assigning algo}. Before assigning every vertex to a partition, the balanced strategy checks the imbalance of computational load between the partitions. Moreover, if all partitions exceed maximum capacity, the algorithm adds a new partition to cater for the upcoming load and thus, to maintain the scalability. We propose a communication-aware balancing strategy and also an adding/removing partitioning technique. These are explained in Section \ref{balancing:strategy} and Section \ref{scalability}  respectively. The partitioning method receives the input from the Graph Loader in a sequential manner. If the algorithm receives an input to add/remove the vertices or edges, it adds/removes the vertices or edges. After removing vertices or edges from a partition, the partitions might become unbalanced. As a result, it is necessary to make the partition balanced. A communication aware balancing strategy is employed here before assigning a vertex to a partition. For the balanced partition we take the number of communications in the balancing method into account. This checks the number of communications over time while balancing the load. The key idea here is to trade-off the number of cut edges with the load imbalance. Algorithm \ref{main-algo} depicts the whole partitioning strategy dynamically in a one-pass manner.    

Where, $v$ is the vertex which has arrived in the stream and Graph Loader decides randomly what kind of input it is and $\alpha$ represent the type of input. The algorithm also receives the edges to remove, which is $edge<v_1, v_2>$. The imbalance parameter is obtained from the calculation of standard deviation of total number of edges in the partitions. Equation \ref{eqn:load-balance} shows the calculation of standard deviation.

$MAXCAP$ is the maximum capacity of a partition and $averageLoad$ is the average edge load of all partitions which can be calculated by $calculateAverageLoad(|V|, P)$ function. The $assignVertex(v,P,V,E(v))$ function is used to assign the vertex to the respective partition the details of the vertex assigning technique is in Section \ref{vertex-assigning-section}. The algorithm deletes the vertices and edges at a point of time.  We use a $updateSummery(partitionIndex,v,MAXCAP,\\averageLoad)$ function to update the graph summary each time we partition or delete any vertices or edges. 

\subsubsection{Vertex Assigning Method}
\label{vertex-assigning-section}
In general, the way to minimise the cut edges between partitions is to allocate the vertices to a partition which contains the greatest number of neighbouring of those vertices. It is always desirable to allocate the connected vertices to the same physical machine. From the stream of graph data, the candidate vertex arrives with its associated edges as shown in Figure~\ref{Stream_fig}. The partitioning algorithm assigns the vertex with its associated edges to a respective machine. Algorithm \ref{Vertex-Assigning algo} shows the vertex allocation strategy. 
SDP algorithm aims to identify the suitable partition of the arrived vertex to minimise the edge-cut. This vertex allocation technique tends to send a vertex to the partition which has the most connected vertices. The algorithm checks all the partitions' information from the partition summary to decide which partition has the most connections of arrived vertices in the stream. The algorithm then allocates the vertex to that particular partition. However, if two or more partitions have the same number of connection of the candidate vertices, the algorithm assigns that candidate vertex to the partition with minimum load. If it does not find a connection in any of the partitions, then the candidate vertex is allocated to the partition randomly in a uniform manner. Finding the partition that has the most connection is calculated with equation \ref{argmax:1} and finding the minimum load of a partition is assessed with Algorithm \ref{finding-minimum-load-algo}.

\begin{equation}
\underset{k \in P}{\operatorname{arg\,max}} \{|E(V_c) \cap P_k| \}
\label{argmax:1}
\end{equation}
Where, $k$ is the number of partitions. $E(v)$ is the associated edges arrived with vertex $v$ and $P(k)$ is the set of vertices in $k$th partition.

\begin{algorithm} [h]
	\begin{flushleft}
		\textbf{INPUT:} $v$= the candidate vertex that available for partition in the stream, $k$ = number of partitions, $E(v)$ is the associated edges arrived with vertex $v$, $P$ = set of partitioned vertices.\\
		\textbf{OUTPUT:} partition index
	\end{flushleft}
	\begin{algorithmic}
		\FOR{$i=0$ to $k$}		
		\STATE $partitionInfoSet <Integer>( P(i))$
		\STATE $intersectSet \gets partitionInfoSet.retainAll(E(v)) $
		
		\STATE $size \gets size of the of the intersectSet$
		\IF {($size > tempSize$)}
		\STATE $tempSize \gets size$
		\STATE $partitionIndex \gets i$
		\STATE $edgeInfoMap[partitionIndex].put(v, E(v))$
		\ELSE
		\IF {($ size == tempSize$)}
		\STATE $paritionIndex \gets i$
		
		\STATE $edgeInfoMap[partitionIndex].put(v, E(v))$
		
		\ELSE
		\STATE $partitionIndex \gets random(k)$
		\STATE $edgeInfoMap[partitionIndex].put(v, E(v))$
		\ENDIF
		\ENDIF
		\ENDFOR 
	\end{algorithmic}
	\caption{Vertex Assigning Method }
	\label{Vertex-Assigning algo}
\end{algorithm}

\begin{algorithm} [H]
	\begin{flushleft}
		\textbf{INPUT:} $partitionInfoMap<p<List>>$, $k$ = number of partitions\\
		\textbf{OUTPUT:} partition index 
	\end{flushleft}
	\begin{algorithmic}
		\STATE $ firstPartitionSize  \gets size of the first partition$
		\FOR{($i =0$ to $k$ )}
		\IF{($firstPartitionSize > ith partition size $)}
		\STATE $ partitionIndex \gets i$
		\ENDIF
		\ENDFOR
	\end{algorithmic}
	\caption{Finding minimum load}	
	\label{finding-minimum-load-algo}
\end{algorithm}

\subsubsection{Communication-aware balancing strategy}
\label{balancing:strategy}
The load of a partition is the number of external and internal connections of that partition. We propose a balancing strategy to keep the partitions as balanced as possible. The number of communications between the partitions is also taken into account in order to decide the imbalance of computational load  among partitions. The number of communications has a great impact on balancing the load. We use average load difference and communication aware load deviation in order to decide the imbalance of the partitions. In this study, we assume that each machine in the distributed system has the same resources and computing power. We used the following variables to complete the load balancing task: The average load difference is $AVG_d $, the threshold is $TH$, and the weighted deviation is $W_{dev}$.

If $AVG_d > TH$ then the algorithm assign the vertex $v$ to partition $P_l$, otherwise the algorithm executes the vertex assigning method (Algorithm \ref{Vertex-Assigning algo}) to assign the vertices to a suitable partition where $v$ is the vertex that has arrived in the stream to be allocated to a partition. The average load difference ($AVG_d$) can be calculated with the following formula :

\begin{equation}
AVG_d =  (P_h -  P_l)/n
\label{avg_d}
\end{equation}

where, $P_h$ is the partition with the highest load, $P_l$ is the partition which has the lowest load and $n$ is the number of partitions. We get the balancing threshold using the following equation: 

\begin{equation}
TH =  W_{dev}- Load_{dev} 
\label{TH}
\end{equation}

where, $Load_{dev}$ is the load deviation among the partitions. The calculation of load the deviation is the Standard Deviation of a load of the partitioning in a distributed system. The weighted deviation leverages the communication with the computational load. Because any partition in a distributed system has large number of communication carry more computation load than other partitions. Weighted deviation decides the imbalance between the partitions. This balancing strategy ensures good computational load distributions. Weighted deviation is denoted by $W_{dev}$ which can be calculated by using the following equation:  

\begin{equation}
W_{dev} = (edge^t/cut^t) * Load_{dev}
\end{equation}

where, $edge^t$  is the edges arrived over time $t$, and $cut^t$  is the cut edges in $t$ time. 

The communication aware balancing strategy ensures a well-balanced computational load while the number of cut edges is also taken into account in deciding the imbalance of a partition.


\subsubsection{Scalability}
\label{scalability}
The performance of the system can be maintained even when the workload reaches to the maximum threshold is using efficient scaling techniques. For efficient scaling both scale out and scale in techniques need to be considered.

Scaling Out: When the capacity of all the partitions exceeds the constraint $C$, then the additional partition needs to be included in the system in order to accommodate the increasing graph data. We use an adding threshold to add a new partition in the system, which can be defined by the following equation: 
\begin{equation}
addingThreshold = \frac{|E^t|}{|P^{t}|}
\end{equation}
where, $|E^t|$ is the total number of edges that has been assigned to all partitions in time $t$ and $|P^{t}|$ is the total number of partitions over time $t$. The threshold decides when to add a new partition in the system. If the $C \leq addingThreshold $, then the system adds a new instance in the system. $C$ is the capacity constraint of a partition which is the maximum computational load of a partition. In this study, we assume that the capacity of all the partitions is the same. 

\textbf{Scaling In:} We use a vertex migration threshold to scale down the resources from the Cloud. The idea is to shut down the unnecessary or unused machine from the system. The determination of shutting down a machine depends on the $l$ value. If two machines have a computational load less than the $l$, the algorithm migrates the vertices and and their associated edges from the source machine to the destination machine. The source machine ($sourceMachine$) is the machine which has the minimum load among the machines; $sourceMachine$ can be defined with Algorithm \ref{finding-minimum-load-algo}. 

\begin{equation}
l = (toleranceParameter * MAXCAP)/100
\end{equation}

Destination machine is the machine which is available to accept more load. We use $destinationThreshold$ to decide the destination partition to migrate the computational load. To determine the availability of the machines to accept more load, we use the  $destinationThreshold$ threshold. A machine accepts computational load until the machine load is less than or equal to the $destinationThreshold$, which keeps some spaces for the upcoming data from the stream.    

\begin{equation}
d = (param * MAXCAP)/100
\end{equation}

\begin{equation}
destinationThreshold = MAXCAP - d	
\end{equation}

\section{Experimental Settings}
\label{experiment-section}
In this section, we discuss the experimental setup, and performance metrics of this study. We use Java programming language to implement both proposed and existing algorithms. Java socket programming is used to implement the distributed environment to partition a graph dynamically. We use Nectar Cloud machines to set up the experimental settings. We use a master machine to allocate the computational load to the worker machines with the partitioning algorithm. Each machine's characteristics are as follows: Ubuntu 18.04 LTS operating system, m2.medium type machine with 30 VCPUs, 6 GB RAM and 30 GB Disk. 

\subsection{Dataset}
We used a variety of synthetic and real graph datasets to evaluate the dynamic partitioning. Table \ref{dataset-table} shows the lists and characteristics of the datasets used in this study.

\begin{table}
	\centering
	\begin{adjustbox}{width=1\linewidth}
		\small
		\begin{tabular}{rlrr}
			\hline
			\textbf{Name of Dataset} & \textbf{$|V|$} & \textbf{$|E|$} & \textbf{Type} \\ 
			\hline
			3elt (Synthetic) \cite{Chris}&4200 &13722 &Finite-element mashes \\ 
			
			GrQc\cite{snapnets}&5242 &14496 &Collaboration Network\\ 
			
			Wiki-vote\cite{snapnets}&	7,115&	99,291&	Social\\	
			
			4elt (Synthetic)\cite{Chris}&	15,606&	45,878&	Finite-element mashes	\\	
			
			AstroPh \cite{snapnets}&	18,772 &198,110	&Citation\\	
			
			Email-enron\cite{snapnets}&	36,692&	183,831&	Communication\\
			Twitter\cite{snapnets}& 81,306&1,768,149& Social\\
			\hline
		\end{tabular}
		
	\end{adjustbox}
	\caption{Characteristics of Datasets}
	\label{dataset-table}
\end{table}

\subsection{Performance Metrics}
We compared our algorithm with the most recent dynamic partitioning algorithm\cite{Abdolrashidi7584916}, 	\cite{7584914}, \cite{Petroni:2015:HSP:2806416.2806424} and \cite{WANG2019804}. We implemented and evaluated both the proposed and baseline algorithms in the similar environment.  The performance of our proposed algorithm is evaluated using the following performance metrics: i) edge-cut ratio; ii) load imbalance; iii) execution time. We observed the number of external connections of a vertex from one partition to another partition as a cut edge. We calculated the ratio of the edge-cut by using the following equation:

\begin{equation}
edge cut ratio =  \frac{|E(u,v)|}{|E|} 
\label{eqn-edge-cut}
\end{equation}
where, $|E| $ is the total number of edges of a graph and $ |E(u,v)|$ the total number of edges between $u$ and $v$ across partitions.

The load of a partition is the number of external and internal connection of a partition. Standard deviation of the total number of external and internal connection(edges) of all the partitions is the load imbalance. The following equation is used to calculate the standard deviation of the total number of external and internal edges of a partition: 
\begin{equation}
load Imbalance=   \sqrt  \frac{\Sigma|e-\bar{e}|^2}{n} 
\label{eqn:load-balance}
\end{equation}

where, $e$ is the total number of external and internal edges of a partition and $n$ is the total number of partitions.   

We measured the execution time from the start of partitioning until the end of partitioning. The time taken to receive the input is also taken into account in the execution time, as the streaming partitioning algorithm executes as the data stream arrives.   

\subsection{Experimental scenario}
In this sub-section we discuss some experimental scenarios of our dynamic partitioning algorithm.  This SDP algorithm takes the stream of vertices and its associated edges as an input in a single-pass manner. This algorithm also accepts the input to remove vertices and edges at a certain point, in order to test the dynamism of our algorithm. 

\subsubsection{Adding/Deleting Vertex}
In a dynamic graph processing system, the addition/deletion of vertices or edges happens over time as per the demand of a graph application. Consequently, the partitioned graph structure changes over time, which creates the unbalanced partitions and also increases the number of communications between partitions. The proposed dynamic graph partitioning algorithm accepts the graph input sequentially. It adds and deletes the vertices dynamically in a streaming manner. The algorithm assigns the vertex to the respective partition as it arrives. 

In a regular interval of time, the algorithm adds and deletes the graph data from the input dataset. In each interval, we add 25\% of the dataset and then delete 5\% of the dataset from a respective partition. In each interval, the algorithm observes edge-cut and the number of partitions used. The number of vertices to add ($nAddVertex$) and delete ($nDelVertex$) in each time interval by using the following formulae is as follows: 

\begin{equation}
nAddVertex = (totVertex*addvertex(\%))/100
\end{equation}

where, $totalVertex$ is the total number vertices of the input dataset, $addvertex(\%)$ and $deletevertex(\%)$ are the percentages of the entire dataset of adding and deleting the vertices respectively. 

\begin{equation}
nDelVertex = (totVertex*deletevertex(\%))/100	
\end{equation}

\subsubsection{Adding a partition dynamically}
Initially, the partitioning starts with a master machine and a worker instance in the Nectar Cloud environment.Number of worker machines is equal to the number of partitions, the number of worker increases or decreases based on load. A capacity constraint $C$ of each worker machine is used to check the maximum capacity of a worker machine. If all the running worker machines have reached the maximum capacity of $C$, the algorithm dynamically creates and launches another instance to accommodate the ever-increasing graph data load. 
Over time $t$, some vertices/edges may be deleted from a worker machine, making the worker machine available to receive more workload. According to the vertex assigning algorithm and balancing strategy, the master machine assigns the vertices to that available worker. Section \ref{scalability} explains in detail the criteria for adding a new partition dynamically.

\subsubsection{Deleting partition dynamically}
As per the demands of the workload, the dynamic algorithm removes the unnecessary/unused instances from the system. Whenever any worker machine has the capacity to receive more load, it accepts the load until it has 5\% capacity available.  

\begin{figure*}
	\centering
	\subfloat[AstroPh Dataset]{%
		\includegraphics[width=.5\textwidth]{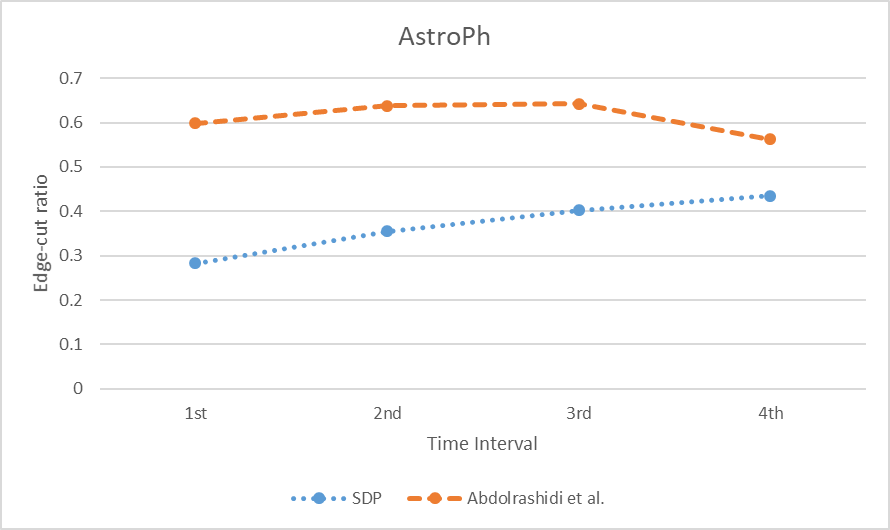}}\hfill
	\subfloat[Copter Dataset]{%
		\includegraphics[width=.5\textwidth]{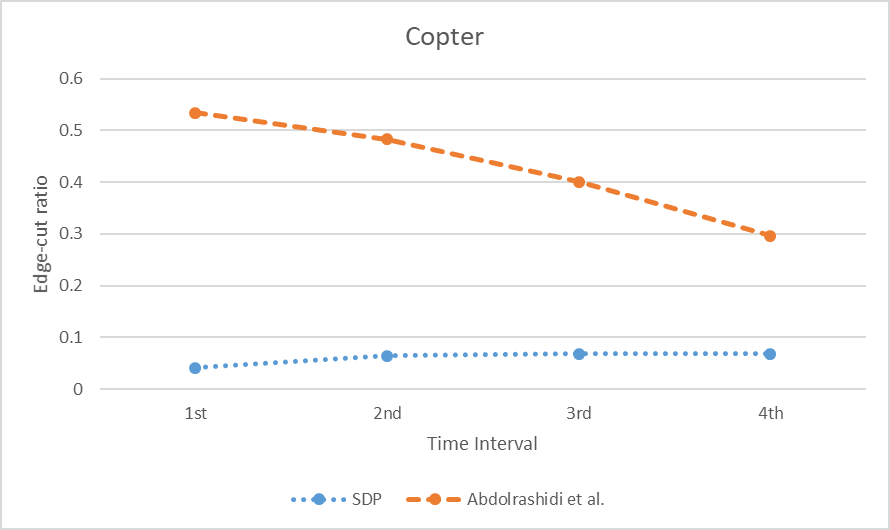}}\hfill
	\subfloat[Email-Enron Dataset]{%
		\includegraphics[width=.5\textwidth]{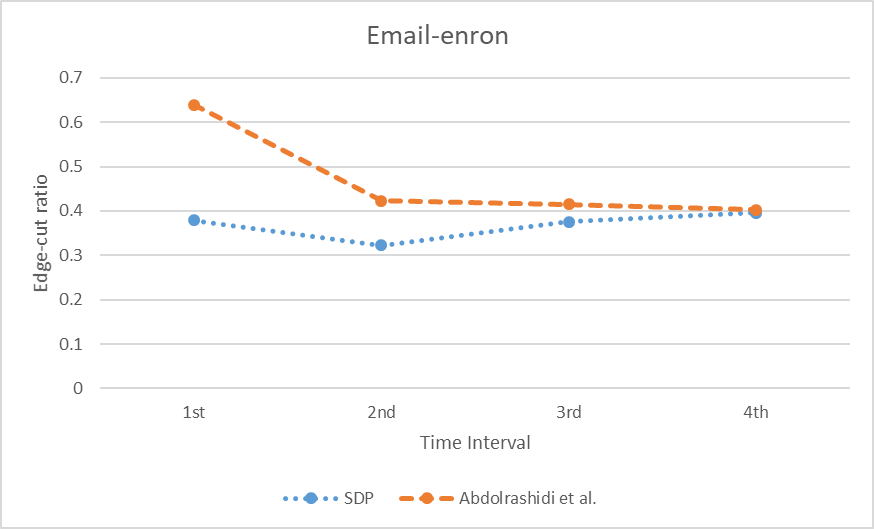}}\hfill
	\subfloat[GrQc Dataset]{%
		\includegraphics[width=.5\textwidth]{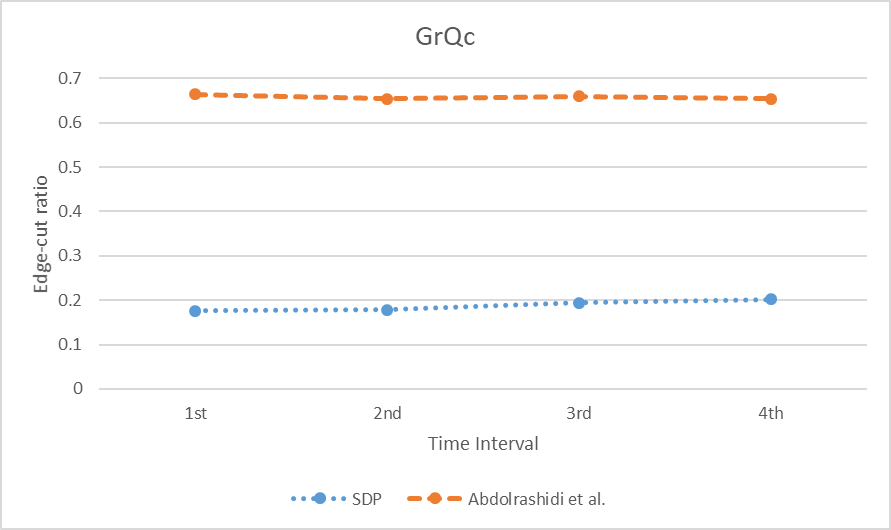}}\hfill
	\caption{Edge-cut comparison of different datasets in different time-interval}
	\label{Edge-cut-Comparison}
\end{figure*}

\section{Result Discussion}
\label{resul-discussion}
In this section, we discuss the results evaluated from the experiments with different types of datasets. The comparison of our algorithm with the existing algorithms is also discussed here.  

\subsection{Edge-cut comparison}  
Edge-cut ratio indicates the performance of graph partitioning in terms of the communication overhead between the partitions. A higher edge-cut indicates higher communication overhead between machines in a distributed system.

We captured the edge-cut over time at an interval of every 25\% of the whole dataset. We compared our algorithm with a recent well-performing, dynamic partitioning algorithm. Figure \ref{Edge-cut-Comparison} illustrates the edge-cut comparison of a number of datasets from different ranges. It is clearly seen that our algorithm obtains a better edge-cut ratio than the existing algorithm. For the Copter and GrQc dataset the algorithm has an 80\%-90\% reduction of edge-cut at the beginning of the partitioning. As shown in Figure \ref{Edge-cut-Comparison}(a-d), the edge-cut ratio goes down when the partitions have more added vertices. It is expected that when the partitions receive more information of a graph, the partitioning performance improves.       

\begin{figure} 
	\includegraphics[width=\linewidth]{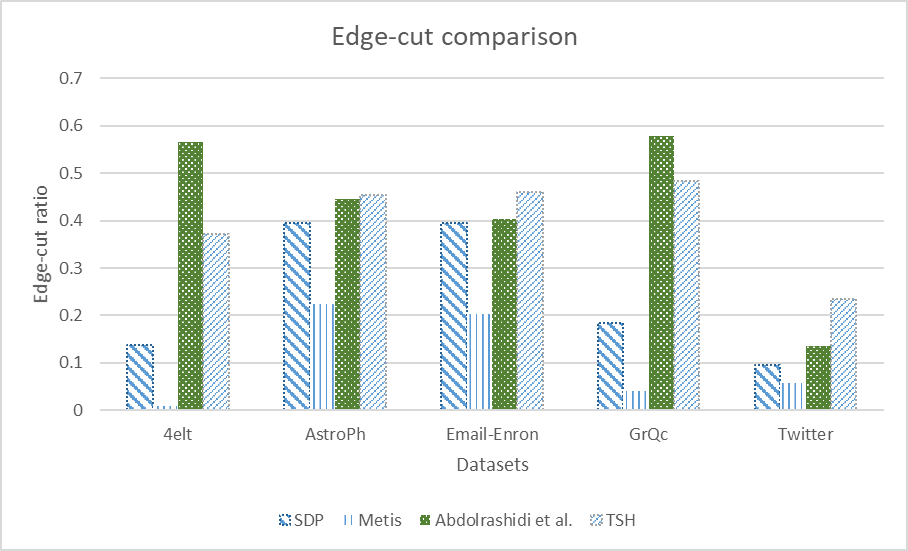}
	\caption{Edge-cut comparison}
	\label{comparison-with-static}
\end{figure}

We also compare our algorithm with the METIS algorithm, which is the state-of-the-art graph partitioning algorithm of all time.  With this comparison, we observe the edge-cut ratio differences between a dynamic and a static graph partitioning algorithm. As shown in Figure \ref{comparison-with-static}, it is obvious that our algorithm performs better for all the datasets than previous algorithms. As expected, METIS algorithm outperformed all other algorithms. This is understandable , as it is difficult for a streaming algorithm to achieve a better edge-cut ratio than an offline graph partitioning. Because, offline graph partitioning algorithm has entire graph information before the start partitioning a graph.

\begin{figure*}
	\centering
	\subfloat[Addition]{%
		\includegraphics[width=.5\textwidth]{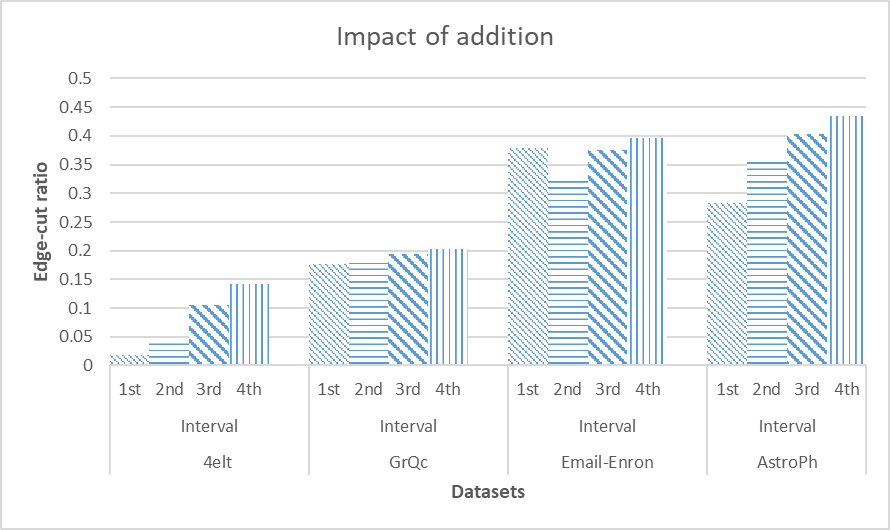}}\hfill
	\subfloat[Deletion]{%
		\includegraphics[width=.5\textwidth]{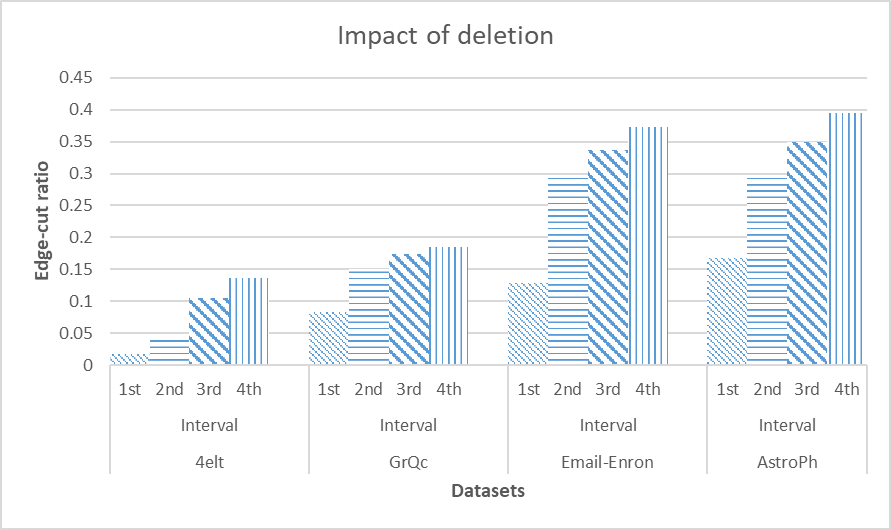}}\hfill
	\caption{Impact of addition/deletion with edge-cut}
	\label{Imact-Addition-Edge-cut-Comparison}
\end{figure*}

\subsection{Impact of addition and deletion}
In this section, we examined the effect of dynamically adding and deleting the vertices or edges for a variety of datasets. Figure \ref{Imact-Addition-Edge-cut-Comparison} shows the trend of the edge-cut ratio over time $t$. In most of the cases, the edge-cut ratio decreases after deletion, while the graph changes in $t$ time. Figure \ref{Imact-Addition-Edge-cut-Comparison} shows the impact of the addition and deletion of vertices and edges over time. We capture the edge-cut performance in four intervals.    

At each interval after deleting vertices and edges from the partitions, the number of edge-cuts decreases as expected. However, as time goes by, the ratio of edge-cuts increases as the deletion percentage is less than the addition percentage. However, an exception happens with the Email-enron dataset, as shown in Figure \ref{Imact-Addition-Edge-cut-Comparison}(a), at the 2nd interval. The edge-cut ratio after the addition is less than the 1st interval. This is because the deleted vertices were connected with a large number of internal edges which creates a slight drop off edge-cut ratio.

\subsection{Load Imbalance Comparison}
\begin{figure} 
	\includegraphics[width=\linewidth]{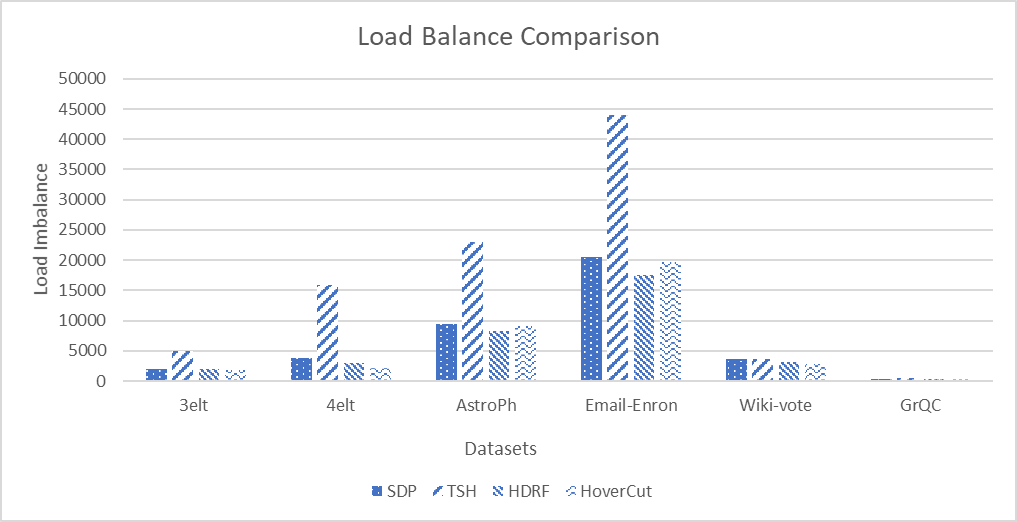}
	\caption{Load Balance Comparison}
	\label{load-balance}
\end{figure}

As shown in Figure \ref{load-balance}, we illustrate the load imbalance comparison between our algorithm and previous algorithms. It is obvious that the reduction of load imbalance in our algorithm is better than the previous algorithms for all the datasets. Our streaming algorithm manages to reduce the 60\% -70\% load imbalance for all the datasets, except the GrQC dataset. The GrQC dataset performed almost similarly to previous algorithms. However, our algorithm performed well in reducing the edge-cut for the GrQC dataset.

\subsection{Impact of number of partitions}
\begin{figure} 
	\includegraphics[width=\linewidth]{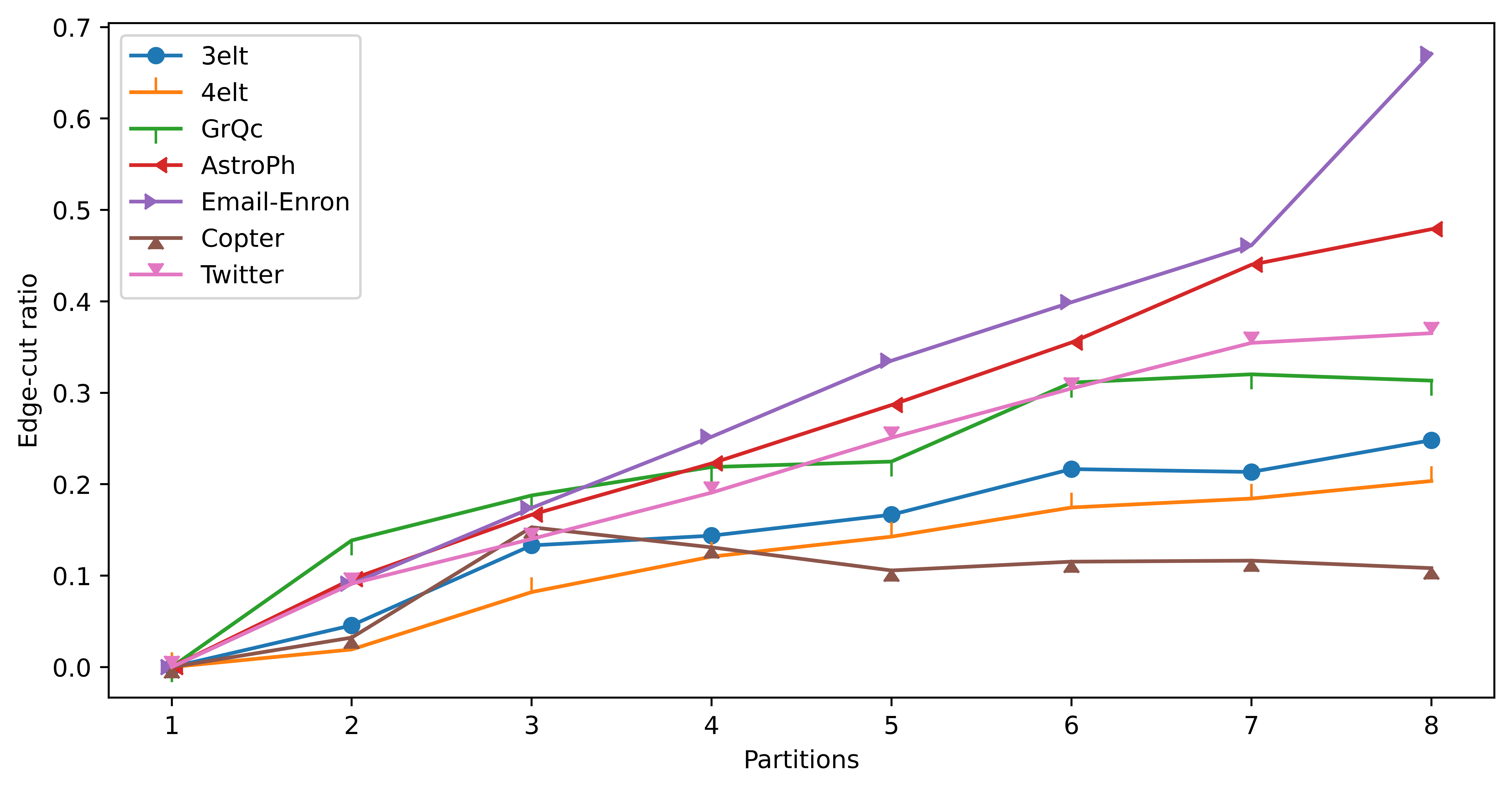}
	\caption{Impact of number of partitions}
	\label{impact-partition}
\end{figure}

This section discusses the effect of the number of partitions in terms of communication cost. As shown in Figure \ref{impact-partition}, it is obvious that the communication cost increases as the number of partition increases. However, in the Copter dataset there was a slight decrease of edge-cut after adding the third partition, as there was deletion of vertices happening at that stage of partitioning.    

\begin{figure}
	\includegraphics[width=\linewidth]{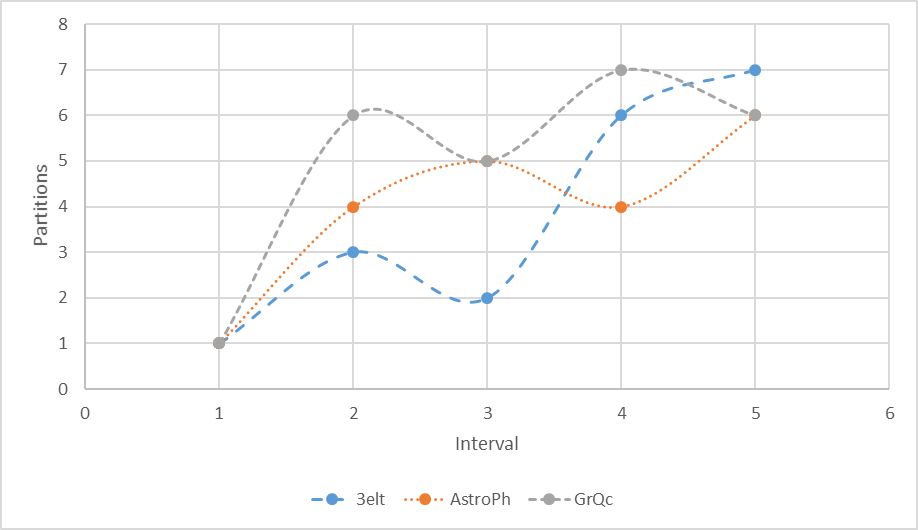}
	\caption{Adding/Removing partitions}
	\label{Add/delete-partition}
\end{figure}

\subsection{Impact of Adding/Removing partitions}
\label{adding}
As shown in Figure \ref{Add/delete-partition}, a number of machines are being added and removed over time for the 3elt, AstroPh, and GrQc datasets. As per the demand of computational load, our partitioning method keeps adding and removing the machines based on the adding/removing criteria. It shows in Figure \ref{Add/delete-partition}, in each interval, the number of worker machine changes because of the dynamic graph updates over time and the demand of worker machine increases and decreases. 

\begin{figure}
	\includegraphics[width=\linewidth]{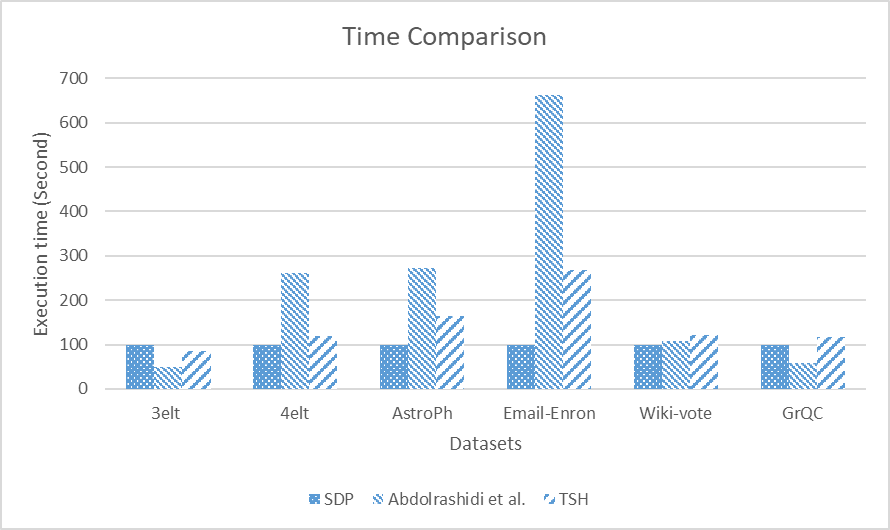}
	\caption{Execution time Comparison}	 
	\label{Time:comparison}
\end{figure}
\setlength\abovedisplayskip{0pt}

\subsection{Time Comparison}
In this section, we discuss the execution time for completing the partitioning task. We calculated the time from the beginning of the algorithm to the end of the execution of a dataset. Figure \ref{Time:comparison} shows the streaming execution time and, it includes the partitioning and input receiving time, because our algorithm does the partitioning task while receiving the input.    

Figure \ref{Time:comparison} shows that our algorithm significantly reduces the execution time over the previous algorithms for most of the datasets except 3elt and GrQC.  

\subsection{Time Complexity}
The time complexity for the SDP algorithm is $O(n+m+w+klogk)$, where $n$ is the number of vertices and $m$ is the number of edges, $w$ is the number of traversals in the stream window for assigning each vertex to a partition, and $k$ is the number of partitions. SDP takes graph input vertex by vertex, where $n$ is the total number of execution to partition an entire graph and $w$ is the number of operations required to traverse through the whole window. Furthermore, $m$ number of operations we need that depends on the number of edges of a vertex has in upcoming vertices.

\section{Conclusion and Future Work}
\label{conclusion}
The major issue in dynamic graph partitioning is to allocate the computational load as it arrives in real time, and to utilise the resources as needed, while, at the same time, minimising the communication and balancing the load as much as possible. This all happens in a real time manner. This study overcomes these issues and proposes a novel dynamic graph partitioning technique. Few studies have been undertaken on the streaming graph partitioning technique with static graph data. However, dealing with a dynamic graph at the same time partitioning in a streaming manner in a Cloud environment, has not yet been studied. This research focuses on partitioning a dynamic graph in a streaming manner. We evaluated our partitioning algorithm with the number of Nectar Cloud instances; every instances was equipped with the same resources. 

In this paper, we propose a partitioning technique of a dynamic graph in a streaming manner. The study demonstrated a substantial improvement in reducing the edge-cut ratio for all the datasets. It also shows an excellent performance 60\%-70\% reduction of the load imbalance in most of the datasets. A communication-aware balancing strategy to balance the computational load among the partitions was suggested. A dynamic auto-scaling method is employed in this study to provision and de-provision the Cloud resources as per the demand of the computational load in a real-time manner. We evaluated the dynamic algorithm in a homogeneous Cloud environment. 

In the future, we will look into the problem of dynamic graph repartitioning in a stream manner. We will also explore the heterogeneous Cloud resources for auto-scaling purposes in streaming graph partitioning.

\bibliographystyle{IEEEtran}
\bibliography{reference}

\begin{thebibliography}{10}
\providecommand{\url}[1]{#1}
\csname url@samestyle\endcsname
\providecommand{\newblock}{\relax}
\providecommand{\bibinfo}[2]{#2}
\providecommand{\BIBentrySTDinterwordspacing}{\spaceskip=0pt\relax}
\providecommand{\BIBentryALTinterwordstretchfactor}{4}
\providecommand{\BIBentryALTinterwordspacing}{\spaceskip=\fontdimen2\font plus
\BIBentryALTinterwordstretchfactor\fontdimen3\font minus
  \fontdimen4\font\relax}
\providecommand{\BIBforeignlanguage}[2]{{%
\expandafter\ifx\csname l@#1\endcsname\relax
\typeout{** WARNING: IEEEtran.bst: No hyphenation pattern has been}%
\typeout{** loaded for the language `#1'. Using the pattern for}%
\typeout{** the default language instead.}%
\else
\language=\csname l@#1\endcsname
\fi
#2}}
\providecommand{\BIBdecl}{\relax}
\BIBdecl

\bibitem{Vaquero:2013:APL:2523616.2525943}
\BIBentryALTinterwordspacing
L.~Vaquero, F.~Cuadrado, D.~Logothetis, and C.~Martella, ``Adaptive
  partitioning for large-scale dynamic graphs,'' in \emph{Proceedings of the
  4th Annual Symposium on Cloud Computing}, ser. SOCC '13.\hskip 1em plus 0.5em
  minus 0.4em\relax New York, NY, USA: ACM, 2013, pp. 35:1--35:2. [Online].
  Available: \url{http://doi.acm.org/10.1145/2523616.2525943}
\BIBentrySTDinterwordspacing

\bibitem{Walshaw:1997:PDG:281659.281661}
\BIBentryALTinterwordspacing
C.~Walshaw, M.~Cross, and M.~Everett, ``Parallel dynamic graph partitioning for
  adaptive unstructured meshes,'' \emph{J. Parallel Distrib. Comput.}, vol.~47,
  no.~2, pp. 102--108, Dec. 1997. [Online]. Available:
  \url{http://dx.doi.org/10.1006/jpdc.1997.1407}
\BIBentrySTDinterwordspacing

\bibitem{6676492}
F.~Rahimian, A.~H. Payberah, S.~Girdzijauskas, M.~Jelasity, and S.~Haridi,
  ``Ja-be-ja: A distributed algorithm for balanced graph partitioning,'' in
  \emph{2013 IEEE 7th International Conference on Self-Adaptive and
  Self-Organizing Systems}, Sept 2013, pp. 51--60.

\bibitem{Stanton2012}
\BIBentryALTinterwordspacing
I.~Stanton and G.~Kliot, ``Streaming graph partitioning for large distributed
  graphs,'' in \emph{Proceedings of the 18th ACM SIGKDD International
  Conference on Knowledge Discovery and Data Mining}, ser. KDD '12.\hskip 1em
  plus 0.5em minus 0.4em\relax New York, NY, USA: ACM, 2012, pp. 1222--1230.
  [Online]. Available: \url{http://doi.acm.org/10.1145/2339530.2339722}
\BIBentrySTDinterwordspacing

\bibitem{Tsourakakis:2014:FSG:2556195.2556213}
\BIBentryALTinterwordspacing
C.~Tsourakakis, C.~Gkantsidis, B.~Radunovic, and M.~Vojnovic, ``Fennel:
  Streaming graph partitioning for massive scale graphs,'' in \emph{Proceedings
  of the 7th ACM International Conference on Web Search and Data Mining}, ser.
  WSDM '14.\hskip 1em plus 0.5em minus 0.4em\relax New York, NY, USA: ACM,
  2014, pp. 333--342. [Online]. Available:
  \url{http://doi.acm.org/10.1145/2556195.2556213}
\BIBentrySTDinterwordspacing

\bibitem{10.1145/3290688.3290711}
\BIBentryALTinterwordspacing
M.~A.~K. Patwary, S.~Garg, and B.~Kang, ``Window-based streaming graph
  partitioning algorithm,'' in \emph{Proceedings of the Australasian Computer
  Science Week Multiconference}, ser. ACSW 2019.\hskip 1em plus 0.5em minus
  0.4em\relax New York, NY, USA: Association for Computing Machinery, 2019.
  [Online]. Available: \url{https://doi.org/10.1145/3290688.3290711}
\BIBentrySTDinterwordspacing

\bibitem{Cheng:2012:KTP:2168836.2168846}
\BIBentryALTinterwordspacing
R.~Cheng, J.~Hong, A.~Kyrola, Y.~Miao, X.~Weng, M.~Wu, F.~Yang, L.~Zhou,
  F.~Zhao, and E.~Chen, ``Kineograph: Taking the pulse of a fast-changing and
  connected world,'' in \emph{Proceedings of the 7th ACM European Conference on
  Computer Systems}, ser. EuroSys '12.\hskip 1em plus 0.5em minus 0.4em\relax
  New York, NY, USA: ACM, 2012, pp. 85--98. [Online]. Available:
  \url{http://doi.acm.org/10.1145/2168836.2168846}
\BIBentrySTDinterwordspacing

\bibitem{Zaharia2010}
\BIBentryALTinterwordspacing
M.~Zaharia, M.~Chowdhury, M.~J. Franklin, S.~Shenker, and I.~Stoica, ``Spark:
  Cluster computing with working sets,'' in \emph{Proceedings of the 2Nd USENIX
  Conference on Hot Topics in Cloud Computing}, ser. HotCloud'10.\hskip 1em
  plus 0.5em minus 0.4em\relax Berkeley, CA, USA: USENIX Association, 2010, pp.
  10--10. [Online]. Available:
  \url{http://dl.acm.org/citation.cfm?id=1863103.1863113}
\BIBentrySTDinterwordspacing

\bibitem{Newman06062006}
\BIBentryALTinterwordspacing
M.~E.~J. Newman, ``Modularity and community structure in networks,''
  \emph{Proceedings of the National Academy of Sciences}, vol. 103, no.~23, pp.
  8577--8582, 2006. [Online]. Available:
  \url{http://www.pnas.org/content/103/23/8577.abstract}
\BIBentrySTDinterwordspacing

\bibitem{Ahmed:2013:DLN:2488388.2488393}
\BIBentryALTinterwordspacing
A.~Ahmed, N.~Shervashidze, S.~Narayanamurthy, V.~Josifovski, and A.~J. Smola,
  ``Distributed large-scale natural graph factorization,'' in \emph{Proceedings
  of the 22Nd International Conference on World Wide Web}, ser. WWW '13.\hskip
  1em plus 0.5em minus 0.4em\relax New York, NY, USA: ACM, 2013, pp. 37--48.
  [Online]. Available: \url{http://doi.acm.org/10.1145/2488388.2488393}
\BIBentrySTDinterwordspacing

\bibitem{DBLP:journals/corr/VaqueroCLM13}
\BIBentryALTinterwordspacing
L.~M. Vaquero, F.~Cuadrado, D.~Logothetis, and C.~Martella, ``xdgp: {A} dynamic
  graph processing system with adaptive partitioning,'' \emph{CoRR}, vol.
  abs/1309.1049, 2013. [Online]. Available:
  \url{http://arxiv.org/abs/1309.1049}
\BIBentrySTDinterwordspacing

\bibitem{Huang:2016:LLE:2904483.2904486}
\BIBentryALTinterwordspacing
J.~Huang and D.~J. Abadi, ``Leopard: Lightweight edge-oriented partitioning and
  replication for dynamic graphs,'' \emph{Proc. VLDB Endow.}, vol.~9, no.~7,
  pp. 540--551, Mar. 2016. [Online]. Available:
  \url{http://dx.doi.org/10.14778/2904483.2904486}
\BIBentrySTDinterwordspacing

\bibitem{Yang:2012:TEP:2213836.2213895}
\BIBentryALTinterwordspacing
S.~Yang, X.~Yan, B.~Zong, and A.~Khan, ``Towards effective partition management
  for large graphs,'' in \emph{Proceedings of the 2012 ACM SIGMOD International
  Conference on Management of Data}, ser. SIGMOD '12.\hskip 1em plus 0.5em
  minus 0.4em\relax New York, NY, USA: ACM, 2012, pp. 517--528. [Online].
  Available: \url{http://doi.acm.org/10.1145/2213836.2213895}
\BIBentrySTDinterwordspacing

\bibitem{Xu:2014:LLD:2733085.2733097}
\BIBentryALTinterwordspacing
N.~Xu, L.~Chen, and B.~Cui, ``Loggp: A log-based dynamic graph partitioning
  method,'' \emph{Proc. VLDB Endow.}, vol.~7, no.~14, pp. 1917--1928, Oct.
  2014. [Online]. Available: \url{http://dx.doi.org/10.14778/2733085.2733097}
\BIBentrySTDinterwordspacing

\bibitem{Duong:2013:SSN:2433396.2433424}
\BIBentryALTinterwordspacing
Q.~Duong, S.~Goel, J.~Hofman, and S.~Vassilvitskii, ``Sharding social
  networks,'' in \emph{Proceedings of the Sixth ACM International Conference on
  Web Search and Data Mining}, ser. WSDM '13.\hskip 1em plus 0.5em minus
  0.4em\relax New York, NY, USA: ACM, 2013, pp. 223--232. [Online]. Available:
  \url{http://doi.acm.org/10.1145/2433396.2433424}
\BIBentrySTDinterwordspacing

\bibitem{6172626}
J.~M. Pujol, V.~Erramilli, G.~Siganos, X.~Yang, N.~Laoutaris, P.~Chhabra, and
  P.~Rodriguez, ``The little engine(s) that could: Scaling online social
  networks,'' \emph{IEEE/ACM Transactions on Networking}, vol.~20, no.~4, pp.
  1162--1175, Aug 2012.

\bibitem{Mondal:2012:MLD:2213836.2213854}
\BIBentryALTinterwordspacing
J.~Mondal and A.~Deshpande, ``Managing large dynamic graphs efficiently,'' in
  \emph{Proceedings of the 2012 ACM SIGMOD International Conference on
  Management of Data}, ser. SIGMOD '12.\hskip 1em plus 0.5em minus 0.4em\relax
  New York, NY, USA: ACM, 2012, pp. 145--156. [Online]. Available:
  \url{http://doi.acm.org/10.1145/2213836.2213854}
\BIBentrySTDinterwordspacing

\bibitem{Abdolrashidi7584916}
A.~Abdolrashidi and L.~Ramaswamy, ``Continual and cost-effective partitioning
  of dynamic graphs for optimizing big graph processing systems,'' in
  \emph{2016 IEEE International Congress on Big Data (BigData Congress)}, 2016,
  pp. 18--25.

\bibitem{WANG2019804}
\BIBentryALTinterwordspacing
N.~Wang, Z.~Wang, Y.~Gu, Y.~Bao, and G.~Yu, ``Tsh: Easy-to-be distributed
  partitioning for large-scale graphs,'' \emph{Future Generation Computer
  Systems}, vol. 101, pp. 804 -- 818, 2019. [Online]. Available:
  \url{http://www.sciencedirect.com/science/article/pii/S0167739X17327723}
\BIBentrySTDinterwordspacing

\bibitem{firth2016workload}
H.~Firth and P.~Missier, ``Workload-aware streaming graph partitioning.''

\bibitem{Riedy:2013:MSD:2425676.2425689}
\BIBentryALTinterwordspacing
J.~Riedy and D.~A. Bader, ``Massive streaming data analytics: A graph-based
  approach,'' \emph{XRDS}, vol.~19, no.~3, pp. 37--43, 2013. [Online].
  Available: \url{http://doi.acm.org/10.1145/2425676.2425689}
\BIBentrySTDinterwordspacing

\bibitem{Tsourakakis:2015:SGP:2817946.2817950}
\BIBentryALTinterwordspacing
C.~Tsourakakis, ``Streaming graph partitioning in the planted partition
  model,'' in \emph{Proceedings of the 2015 ACM on Conference on Online Social
  Networks}, ser. COSN '15.\hskip 1em plus 0.5em minus 0.4em\relax New York,
  NY, USA: ACM, 2015, pp. 27--35. [Online]. Available:
  \url{http://doi.acm.org/10.1145/2817946.2817950}
\BIBentrySTDinterwordspacing

\bibitem{Petroni:2015:HSP:2806416.2806424}
\BIBentryALTinterwordspacing
F.~Petroni, L.~Querzoni, K.~Daudjee, S.~Kamali, and G.~Iacoboni, ``Hdrf:
  Stream-based partitioning for power-law graphs,'' in \emph{Proceedings of the
  24th ACM International on Conference on Information and Knowledge
  Management}, ser. CIKM '15.\hskip 1em plus 0.5em minus 0.4em\relax New York,
  NY, USA: ACM, 2015, pp. 243--252. [Online]. Available:
  \url{http://doi.acm.org/10.1145/2806416.2806424}
\BIBentrySTDinterwordspacing

\bibitem{7584914}
H.~P. Sajjad, A.~H. Payberah, F.~Rahimian, V.~Vlassov, and S.~Haridi,
  ``Boosting vertex-cut partitioning for streaming graphs,'' in \emph{2016 IEEE
  International Congress on Big Data (BigData Congress)}, June 2016, pp. 1--8.

\bibitem{Chris}
C.~Walshaw, ``The graph partitioning archive,''
  \url{http://http://chriswalshaw.co.uk/partition/}, 2016.

\bibitem{snapnets}
J.~Leskovec and A.~Krevl, ``{SNAP Datasets}: {Stanford} large network dataset
  collection,'' \url{http://snap.stanford.edu/data}, 2014.

\end{thebibliography}
%

\begin{IEEEbiography}[{\includegraphics[width=1in,height=1.25in,clip,keepaspectratio]{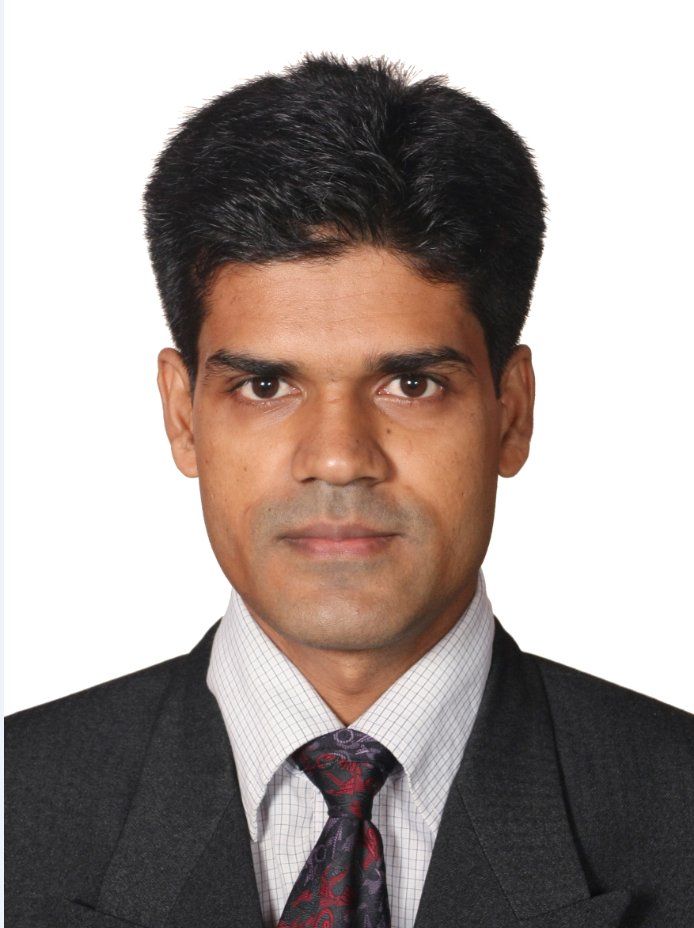}}]{Dr. Md Anwarul Kaium Patwary}
 completed his PhD in Computer Engineering from the University of Tasmania in 2020. Currently, he is working as an Associate Lecturer at The University of Western Australia. His research interests include dynamic graph partitioning, graph machine learning, deep learning, and distributed computing.
\end{IEEEbiography}

\begin{IEEEbiography}[{\includegraphics[width=1in,height=1.25in,clip,keepaspectratio]{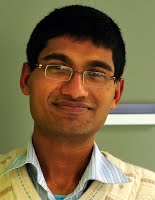}}]{Dr. Saurabh Garg}
  is currently a Lecturer with the University of Tasmania, Australia. He is one of the few Ph.D. students who completed in less than three years from the University of Melbourne. He has authored over 40 papers in highly cited journals and conferences. During his Ph.D., he has been received various special scholarships for his Ph.D. candidature. His research interests include resource management, scheduling, utility and grid computing, Cloud computing, green computing, wireless networks, and ad hoc networks.
\end{IEEEbiography}

\begin{IEEEbiography}[{\includegraphics[width=1in,height=1.25in,clip,keepaspectratio]{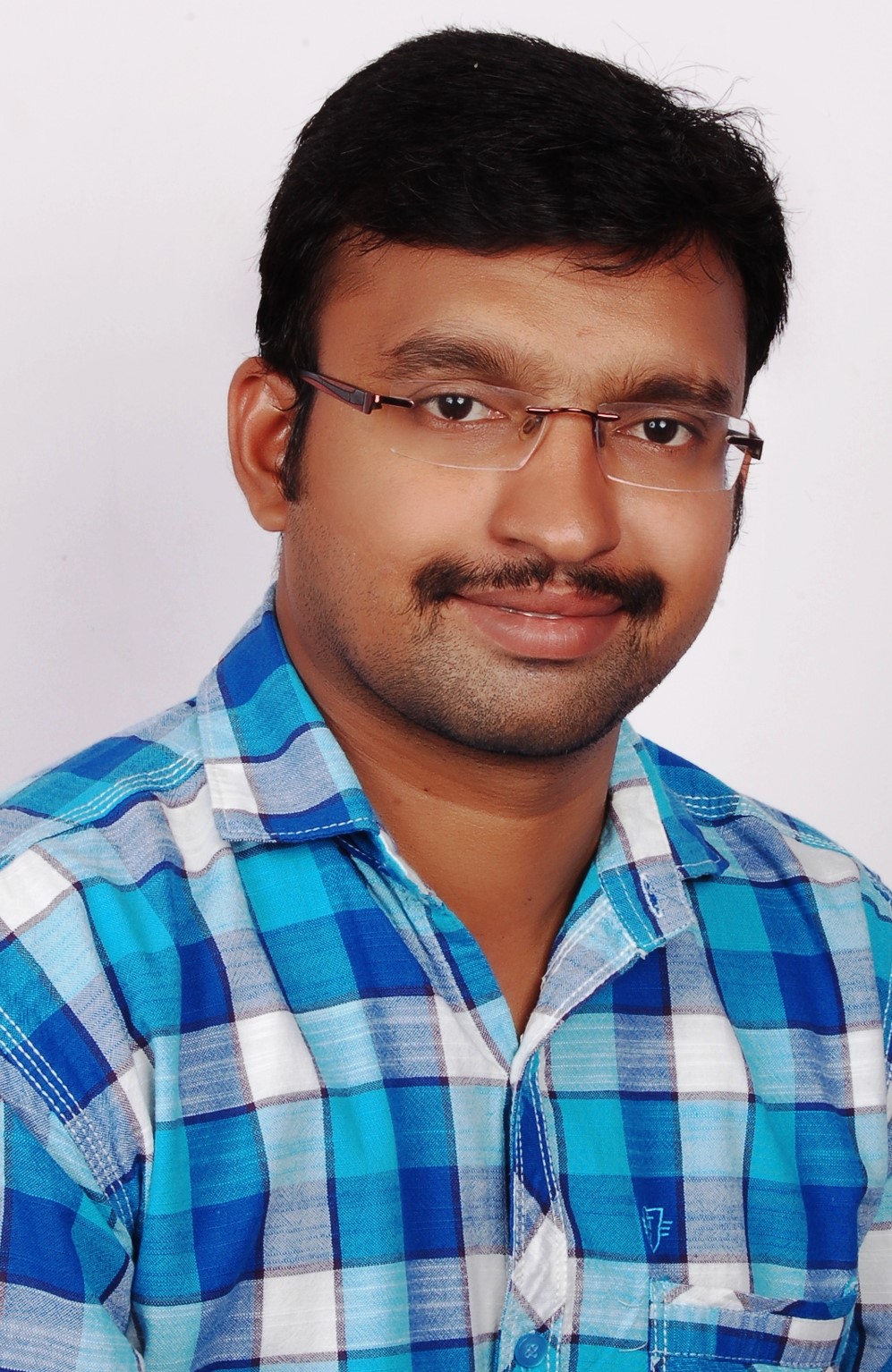}}]{ Sudheer Kumar Battula }
 received his Master of Technology degree in software
engineering in 2012. He is currently pursuing his Ph.D. studies on resource management in Fog computing environment with the University of
Tasmania. He has been awarded Tasmania Graduate Research Scholarship (TGRS) for supporting his studies. His research interests includes Fog computing, Blockchain, Distributed file systems, Cloud computing, Internet of Things (IoT), and Big Data.
\end{IEEEbiography}

\begin{IEEEbiography}[{\includegraphics[width=1in,height=1.25in,clip,keepaspectratio]{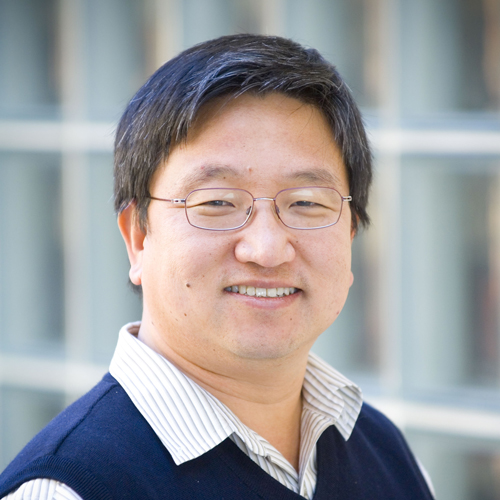}}]{  Byeong Kang }
 
 is a Professor in School of Technology, Environments and Design, University  of  Tasmania,  Australia.  Prof. Kang  received  his  PhD  from  the University of New South Wales, Sydney, in 1996, and has worked as a visiting  researcher  in  the  Advanced Research Lab HITACHI situated in Japan. His  research  interests  include  basic knowledge   acquisition   methods   and applied research in Internet systems as well as medical expert systems. Pro. Kang leads the Smart Services and Systems research group of postdoctoral scientists, which has carried out fundamental and applied research in research areas, expert systems, SNS analysis and smart industry areas. He has served as  a  chair  and  steering  committee  member  in  many international organizations and during conferences.
\end{IEEEbiography}

\end{document}